\documentclass{llncs}

\usepackage[T1]{fontenc}
\usepackage[utf8]{inputenc}
\usepackage{microtype}
\usepackage{amsmath,amssymb}
\usepackage{array}
\usepackage{booktabs}
\usepackage{graphicx}
\usepackage{xcolor}
\usepackage{tikz}
\usetikzlibrary{shapes.geometric,arrows.meta,positioning,fit,backgrounds,calc}
\usepackage{pgfplots}
\pgfplotsset{compat=1.18}
\usepackage{hyperref}
\hypersetup{colorlinks=true,linkcolor=black,citecolor=black,urlcolor=black}
\usepackage{cleveref}
\usepackage{algorithm}
\usepackage{algpseudocode}
\usepackage{multirow}

\newcommand{\sys}{\textnormal{\textsc{Tessera}}}
\newcommand{\BW}{BW_{\mathrm{mem}}}

\newcommand{\Tline}{T_{\mathrm{line}}}

\title{\sys{}: Secure, Near-Line-Rate Weight Streaming for UMA Edge Accelerators}

\author{Animan Naskar}
\authorrunning{A. Naskar}
\institute{Indian Institute of Technology (IIT) Ropar\\
\email{2022csb1297@iitrpr.ac.in}}
\begin{document}
\maketitle



\begin{abstract}
Deploying proprietary Deep Neural Networks (DNNs) on commodity edge devices demands hardware-backed Digital Rights Management (DRM) capable of withstanding both software-level and physical adversaries. In Unified Memory Architecture (UMA) systems, the host CPU and Neural Processing Unit (NPU) share physical DRAM, leaving plaintext model weights directly readable by a compromised OS kernel. Existing defenses fail in this constrained setting: trusted execution environments monopolize scarce memory with permanently reserved regions, while full-memory encryption operates at page granularity. This forces the system to fetch massive 4\,KB memory pages for sub-page tensor tiles, severely crippling bandwidth.

We present \sys{}, a reference architecture for inline, cache-line-granularity weight decryption on UMA edge accelerators. The design intercepts 64-byte AXI bursts, computing AES-256-CTR keystreams in parallel with DRAM fetches. This streams plaintext directly into isolated NPU SRAM, creating a transient memory footprint confined to the active tile and eliminating the need for permanent memory carve-outs. Measurements across three distinct SoC platforms demonstrate that this parallelization hides cryptographic latency behind standard DRAM fetch times, a condition that holds even under worst-case timing variations. Consequently, \sys{} is projected to achieve 98.4\% of the theoretical memory bandwidth ceiling (a mere 1.6\% overhead). Across standard vision and language models, page-level memory encryption suffers up to a $32\times$ bandwidth penalty, whereas \sys{} maintains an optimal $1\times$ footprint for all layer geometries. Finally, \sys{} neutralizes major UMA-specific attack vectors---including physical DRAM extraction, rogue DMA, and compute hijacking---and formally prevents plaintext leakage across sparse tensors.
\end{abstract}

\keywords{Edge AI \and Hardware Security \and Unified Memory Architecture
\and Inline Crypto Engine \and AES-CTR \and DRM \and Neural Processing Unit
\and Cache-Line Encryption}

\section{Introduction}
\label{sec:intro}
Modern edge inference platforms ship with NPUs capable of tens of TOPS,
enabling DNN deployment in applications ranging from autonomous vehicles
to medical diagnostics. The commercial value of underlying model weights
is substantial: manufacturers increasingly treat the model binary as a
licensable asset subject to per-device or per-inference DRM.

Weight protection is fundamentally harder on edge devices than in cloud
settings for three primary reasons. \emph{First}, edge devices are physically
accessible; cold-boot attacks and DRAM interposer probing are well-documented
on consumer hardware~\cite{halderman2008lest,tatar2018throwhammer}.
\emph{Second}, the OS attack surface is far larger on resource-constrained
IoT devices~\cite{mo2021ppfl}. \emph{Third}, the UMA topology prevalent in
mobile SoCs---exemplified by ARM Mali and NVIDIA Jetson---places the CPU
DRAM bus on a path that physically intersects the accelerator's tensor-tile
fetch stream, leaving any plaintext in shared DRAM trivially accessible to a
kernel-level attacker.

These constraints undermine mainstream defenses. Full-DRAM encryption
(AMD SME~\cite{amdSME}, Intel TME~\cite{intelTME}) operates at page
granularity, causing massive traffic amplification for sub-page tensor tiles.
CPU-hosted TEEs (SGX~\cite{costan2016intel}, TrustZone~\cite{armTZ})
introduce data-copy overheads that are difficult to hide in real-time edge
pipelines. Furthermore, systems that rely on statically reserved
protected-memory regions (PVM-style enclaves) permanently reduce the
usable system RAM available to the OS and co-located workloads.

\textbf{This paper.} \sys{} is a vendor-agnostic reference architecture
that addresses these UMA vulnerabilities by confining decryption to a
dedicated Inline Crypto Engine (ICE) physically interposed on the AXI bus.
It decrypts weights just-in-time directly into the NPU's on-chip SRAM,
ensuring plaintext exists only transiently without carving out protected
regions of main memory. Our contributions are:

\begin{itemize}
  \item \textbf{Cache-line-granularity ICE for UMA NPUs
        (\cref{sec:architecture}):} A concrete reference architecture
        for 64-byte AXI-interposed decryption on UMA SoCs, featuring a
        consistent RSA-based key hierarchy, address-derived AES-CTR
        counters, and SMMU stream-ID isolation.

  \item \textbf{Empirical demonstration of full weight extraction
        (\cref{sec:eval-threat}):} A cross-model demonstration (ResNet-18,
        MobileNetV2, DistilBERT-Tiny) proving that 100\% of DNN weights
        are trivially recoverable on a Jetson AGX Xavier via
        \texttt{/dev/mem} with zero exploit, directly validating the UMA
        threat model.

  \item \textbf{Formal necessity of address-derived counters
        (\cref{sec:primitive}):} A proof that any fixed-counter scheme
        leaks plaintext via XOR cancellation over sparse DNN tensors,
        establishing address-derived derivation as a strict cryptographic
        requirement for cache-line DRM.

  \item \textbf{Near-line-rate throughput and optimized energy efficiency
        (\cref{sec:eval}):} Hardware measurements confirm \sys{} safely
        hides cryptographic overhead behind standard DRAM fetch times,
        projecting a throughput of 22.1\,GB/s (98.4\% of the DDR5-4800
        peak ceiling). Unlike page-level defenses that force massive 4\,KB
        fetches---imposing up to a $32\times$ bandwidth penalty---\sys{}
        secures data at the hardware's native 64-byte granularity. This
        eliminates structural traffic amplification and saves over 22\,mJ
        of DRAM PHY energy per inference.
\end{itemize}

\section{Background and Motivation}
\label{sec:background}

\subsection{Unified Memory in Edge Accelerators}

Unlike discrete GPU configurations with separate PCIe-attached DRAM,
UMA SoCs instantiate a single physical DRAM array shared by all on-chip
masters: CPU clusters, NPU, image signal processor, and DMA engines.
The NPU reads tensor tiles via large sequential DMA bursts, bypassing
the CPU L3 cache hierarchy entirely. A consequence is that CPU-side
MMU protections do not prevent a DMA engine from issuing reads to the
same physical pages; isolation additionally requires SMMU stream-ID
firewall rules.

\subsection{Limitations of Page-Level Encryption}

For a tensor tile of $t$ bytes where $t \ll 4096$, decrypting a full
page amplifies effective DRAM traffic by $\lceil 4096/t \rceil$.
This reaches $32\times$ for batch-normalisation parameter fetches
($t = 128$\,B) and $15\times$ for depth-wise convolution weight tiles
($t = 288$\,B). Moreover, page-level ICEs expose decrypted plaintext in
the CPU-accessible memory fabric---a concession a kernel-privileged
attacker can exploit via straightforward \texttt{/dev/mem} reads on
UMA devices.

\subsection{Limitations of Reserved Secure Memory}

A common alternative to shared-DRAM protection is to execute sensitive
workloads inside a statically reserved protected-memory region, such as a
TEE-backed carve-out or PVM-style secure buffer. This approach improves
confidentiality, but it creates a direct capacity trade-off: the reserved
region is no longer available to the host OS, so usable system memory is
reduced by exactly the amount carved out for security. On memory-constrained
edge devices, that trade-off is often expensive; for example, reserving
2\,GB of protected memory on an 8\,GB phone leaves only 6\,GB for normal
applications and services. In addition, the secure region must be sized for
worst-case model footprints even though inference touches only a small tile at
a time, so much of the reserved memory sits idle during execution.

\sys{} avoids this penalty by keeping plaintext out of shared DRAM entirely
and materializing it only just in time inside the NPU's on-chip SRAM. The
result is a small transient footprint rather than a permanently reserved
secure-memory carve-out.

\subsection{Threat Model}
\label{sec:threat}

\begin{definition}[Adversary Capabilities]
  We assume a powerful adversary with full control over the host OS and physical access to off-chip components. Specifically, the adversary possesses: 
  (1) \textbf{System Compromise:} Root/kernel execution on the host CPU, including full IOMMU page-table write access; 
  (2) \textbf{Rogue DMA:} The ability to issue arbitrary DMA transactions from malicious peripherals to any SMMU-unprotected address; 
  (3) \textbf{Physical DRAM Access:} PCB access enabling cold-boot extraction~\cite{halderman2008lest} or DRAM bus interposer probing; and 
  (4) \textbf{Traffic Manipulation:} The capability to replay, reorder, or alias ciphertext blocks presented to the ICE.
\end{definition}

\begin{definition}[Trust Boundary]
  The physical SoC die boundary is trusted. Consistent with standard hardware security models (e.g., ARM TrustZone~\cite{armTZ}, AMD PSP~\cite{amdPSP}), we explicitly exclude invasive silicon attacks (e.g., decapping), fault injection (e.g., glitching), and OEM supply-chain tampering. Physical side-channel analysis of the ICE is likewise out of scope; architectural mitigations for these are discussed in \cref{sec:limitations}.
\end{definition}

\section{The \sys{} Reference Architecture}
\label{sec:architecture}
\begin{figure}[H]
\centering
\includegraphics[width=\linewidth]{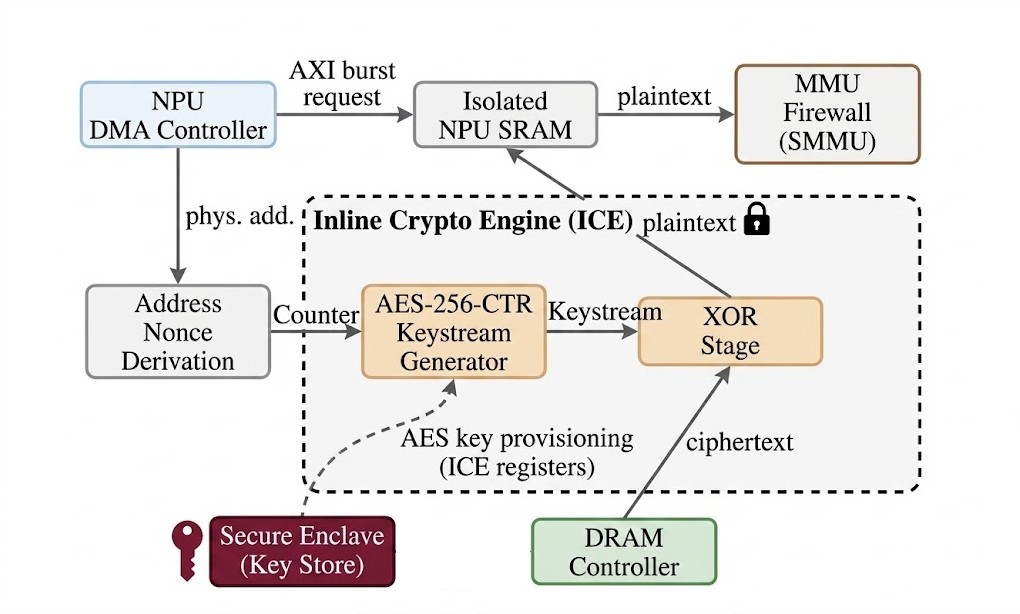}
\caption{\sys{} hardware pipeline. The ICE sits on the path between the
DRAM controller and the NPU DMA fabric. The secure enclave provisions the
AES session key directly into ICE registers; address-derived counters enable
parallel keystream generation; and the SMMU firewall blocks host access to
plaintext SRAM.}
\label{fig:pipeline}
\end{figure}

\Cref{fig:pipeline} summarises the \sys{} datapath and trust boundary.

\subsection{Hardware Root of Trust and Key Hierarchy}
\label{sec:keyhierarchy}

\sys{} relies on a three-layer key hierarchy anchored by a hardware root of trust, utilizing RSA with OAEP padding (RSAES-OAEP~\cite{rfc8017}) for secure key delivery.

\noindent\textbf{Layer~1 --- Device Identity Keypair (DIK).}
During device provisioning, the OEM generates (or injects) a unique
2048-bit (or 4096-bit) RSA keypair $(\mathit{pk}_\mathrm{dev},\, \mathit{sk}_\mathrm{dev})$.
The public key $\mathit{pk}_\mathrm{dev}$ is embedded in a device certificate,
while the private key $\mathit{sk}_\mathrm{dev}$ is fused into on-die eFuses
and accessible only to a hardware Secure Enclave via a restricted interface.
The key has no software-visible interface and is never exposed to shared DRAM.

\noindent\textbf{Layer~2 --- Per-Model Session Key (MSK).}
For each model, the developer generates a random 256-bit AES key
($k_\mathrm{msk}$) and encrypts weights using AES-256-CTR. The key is
provisioned via a bound blob:
\begin{equation}
\label{eq:keyblob}
  \mathcal{B} = \mathsf{RSAES\text{-}OAEP}\!\left(\mathit{pk}_\mathrm{dev},\;
  k_\mathrm{msk} \,\|\, H_\mathit{app}\right)
\end{equation}
where $H_\mathit{app} = \mathsf{SHA\text{-}256}(\text{code-signing certificate})$.
Encryption under $\mathit{pk}_\mathrm{dev}$ ensures \emph{device binding}, while
including $H_\mathit{app}$ enables \emph{application binding}, enforced by the
Secure Enclave at load time. Any modification causes OAEP decoding to fail with
overwhelming probability.

\noindent\textbf{Layer~3 --- ICE Register Provisioning.}
At load time, the untrusted OS supplies $\mathcal{B}$ to the Secure Enclave,
which decrypts it using $\mathit{sk}_\mathrm{dev}$ to recover
$k_\mathrm{msk} \,\|\, H_\mathit{app}$. The Enclave verifies the caller’s
code-signing identity (via secure boot/attestation) against $H_\mathit{app}$.
If valid, it provisions $k_\mathrm{msk}$ into ICE registers via a secure on-die
bus inaccessible to CPU and DMA; otherwise, the load aborts. The session key is
never written to shared DRAM.

\subsection{Cryptographic Primitive: Cache-Line AES-CTR}
\label{sec:primitive}
Out-of-order NPU DMA fetches preclude sequential modes (e.g., CBC, CFB).
\sys{} therefore uses AES-CTR with address-derived counters. Let $L = 64$
bytes, $\mathit{IV}_\mathrm{base}$ be a 96-bit per-model nonce stored in
$\mathcal{B}$, and $P_\mathrm{addr}$ be a cache-line-aligned physical address.
The counter is defined as:
\begin{equation}
  \mathit{CTR}(P_\mathrm{addr}) = \mathit{IV}_\mathrm{base}
    \;\|\; \left\lfloor \frac{P_\mathrm{addr}}{L} \right\rfloor
  \label{eq:ctr}
\end{equation}
forming a 128-bit AES input from the nonce and a 32-bit block index.
To prevent keystream reuse across models or executions with overlapping
address spaces, $\mathit{IV}_\mathrm{base}$ provides per-model domain separation.

\begin{proposition}[Counter Uniqueness]
If $\left\lfloor P_\mathrm{addr}^{(1)}/L \right\rfloor \ne
\left\lfloor P_\mathrm{addr}^{(2)}/L \right\rfloor$ then \\
$\mathit{CTR}(P_\mathrm{addr}^{(1)}) \ne
\mathit{CTR}(P_\mathrm{addr}^{(2)})$, for weight blobs up to
$2^{32} \times L = 256$\,GiB (i.e., without counter wraparound).
\end{proposition}

\begin{proposition}[Address-Aliasing Resistance]
\label{prop:alias}
Remapping physical pages to present the same ciphertext at address
$P' \ne P$ causes the ICE to apply $\mathit{CTR}(P') \ne \mathit{CTR}(P)$,
yielding incorrect plaintext with overwhelming probability. This neutralises
adversary capability~(4) from \cref{sec:threat}.
\end{proposition}

\subsection{The ICE Execution Pipeline}
\label{sec:pipeline}

\Cref{alg:ice} formalises the per-cache-line ICE operation.

\noindent\textbf{Stage~1: DMA Intercept.} The ICE captures each
cache-line-granularity read request on the AXI bus and extracts the
target physical address $P_\mathrm{addr}$.

\noindent\textbf{Stage~2: Parallel Keystream and Fetch.} The ICE
computes $\mathit{CTR}(P)$ per \cref{eq:ctr} in one cycle, then
initiates AES-256 forward encryption. A pipelined AES-256 core
with $R$ pipeline stages produces the keystream in $T_\mathit{ks} = R$
cycles. Concurrently, the DRAM fetch traverses the memory controller
and DDR PHY. Because $T_\mathit{DRAM} \gg T_\mathit{ks}$ on all
evaluated platforms (\cref{sec:eval-latency}), keystream
completion occurs before ciphertext arrival in the common case.

\noindent\textbf{Stage~3: Line-Rate XOR.} The ICE XORs the
64-byte burst combinatorially. The XOR stage is pipelined into
AXI datapath, adding only 1-2 cycles of delay.

\noindent\textbf{Stage~4: Isolated SRAM Write.} Plaintext is placed
on the AXI write channel to the NPU's L2 SRAM, protected by the SMMU
stream-ID firewall (\cref{sec:isolation}).

\begin{algorithm}[H]
  \caption{ICE Per-Cache-Line Decrypt (\sys{})}
  \label{alg:ice}
  \begin{algorithmic}[1]
    \Require Physical address $P_\mathrm{addr}$, ciphertext $C[0..L{-}1]$,
             key $k_\mathrm{msk}$, base nonce $\mathit{IV}_\mathrm{base}$
    \Ensure  Plaintext $M[0..L{-}1]$ written to isolated SRAM
    \State $\mathit{ctr} \gets \mathit{IV}_\mathrm{base} \| \lfloor P/L \rfloor$
           \Comment{1 cycle; address arithmetic}
    \State $\mathit{KS} \gets \mathsf{AES\text{-}256}(k_\mathrm{msk},\; \mathit{ctr})$
           \Comment{parallel with DRAM fetch}
    \State \textbf{await} DRAM burst $C$ on AXI bus
    \State $M \gets C \oplus \mathit{KS}$
           \Comment{bitwise XOR, 1--2 cycles}
    \State \textbf{write} $M$ to isolated NPU SRAM
  \end{algorithmic}
\end{algorithm}

\subsection{Enforcing Unified Memory Isolation}
\label{sec:isolation}

\noindent\textbf{SMMU Stream-ID Firewalling.}
The SoC System Memory Management Unit (SMMU) is configured in TrustZone
Secure World to assign the NPU DMA engine a dedicated stream ID that
maps exclusively to the protected SRAM range. Any CPU-initiated access
to this range results in a bus abort (\texttt{SLVERR}/\texttt{DECERR}
on AXI). The SMMU configuration registers are TrustZone-locked and are
not writable by Normal World software.

\noindent\textbf{SMMU Trust Assumptions.}
\sys{} assumes correct SMMU configuration by trusted firmware and the
absence of implementation flaws. Prior work has demonstrated that
misconfiguration or firmware vulnerabilities (e.g.,~\cite{markettos2019thunderclap})
can undermine IOMMU-based isolation. Thus, SMMU enforcement is not
inherently fail-safe and must be treated as part of the trusted computing
base.

\noindent\textbf{Hardware Tag Propagation.}
For defense in depth, \sys{} leverages hardware tagging where available
(e.g., ARM MTE or equivalent coloring mechanisms). The ICE annotates
plaintext transactions with a restricted tag that is enforced by the
interconnect on subsequent accesses, preventing unauthorized agents from
dereferencing protected data even in the presence of SMMU misconfiguration.

\section{Preemption and Context-Switch Safety}
\label{sec:preemption}

In multi-process environments, OS preemption can interrupt the NPU
mid-inference. Without safeguards, decrypted weights may persist in SRAM
and be exposed to subsequent contexts. \sys{} prevents such leakage via
the hardware preemption hook in \cref{alg:ctx}.
\begin{algorithm}[H]
  \caption{\sys{} Preemption Hook (Revised)}
  \label{alg:ctx}
  \begin{algorithmic}[1]
    \Require Preemption signal $\sigma$
    \Ensure No plaintext persists across the context boundary
    \State Stop issuing new DMA requests from NPU
    \State Drain in-flight AXI transactions
    \State Zero-fill / invalidate plaintext SRAM via hardware scrub engine
    \State Clear ICE key registers
    \State OS proceeds with context switch
    \Statex \hrule\vspace{2pt}\textbf{On resume:}
    \State Re-provision $k_\mathrm{msk}$ into ICE from Secure Enclave
    \State Restart execution from last tile boundary (re-fetch from DRAM)
  \end{algorithmic}
\end{algorithm}

\noindent\textbf{Hardware-Enforced Execution.}
To tolerate a compromised OS, SRAM zero-fill (Step~3) is performed by a
dedicated hardware scrub engine within the ICE datapath, triggered via a
privileged control path inaccessible to the CPU. This ensures that plaintext
removal cannot be bypassed or delayed by software.

\noindent\textbf{Preemption Latency.}
The dominant cost is SRAM zero-fill:
\begin{equation}
  T_\mathit{preempt} = \frac{S_\mathit{SRAM}}{\mathit{BW}_\mathit{SRAM}} + T_\mathit{save}
\end{equation}
where $T_\mathit{save} \approx 1.5\,\mu$s includes state save and
TrustZone switching.

As shown in \cref{tab:preempt}, $T_\mathit{preempt}$ is well below typical
OS context-switch latency ($\geq$100\,$\mu$s). For example, a 2\,MB SRAM
at 512\,GB/s yields $T_\mathit{preempt} \approx 5.4\,\mu$s, indicating
negligible scheduling impact.

\begin{table}[H]
\centering
\footnotesize
\setlength{\tabcolsep}{4pt}
\caption{Instantiated $T_\mathit{preempt}$ for evaluated NPU configurations. Formula: $T_\mathit{preempt} = S_\mathit{SRAM}/\mathit{BW}_\mathit{SRAM} + T_\mathit{save}$ ($T_\mathit{save}=1.5$\,$\mu$s).}
\label{tab:preempt}
\begin{tabular}{@{}lccr@{}}
\toprule
\textbf{Platform (NPU)} & $S_\mathit{SRAM}$ & $\mathit{BW}_\mathit{SRAM}$ &
  $T_\mathit{preempt}$ \\
\midrule
Intel i9-12900H (iGPU L2\$) & 2\,MB & 512\,GB/s &  5.4\,$\mu$s \\
Jetson AGX Xavier (DLA)     & 4\,MB & 480\,GB/s &  9.8\,$\mu$s \\
Jetson AGX Orin (DLA)       & 4\,MB & 960\,GB/s &  5.7\,$\mu$s \\
\midrule
\multicolumn{3}{@{}l}{\footnotesize OS scheduler context-switch floor} &
  $\geq$100\,$\mu$s \\
\bottomrule
\end{tabular}
\end{table}
\section{Evaluation}
\label{sec:eval}

\subsection{Methodology and Scope}
\label{sec:eval-methodology}

As \sys{} is a reference architecture pending silicon fabrication, we evaluate it using a combination of empirical measurements on proxy hardware and analytical modeling. Our methodology is designed to establish feasibility and bounding characteristics along four axes: (i) UMA vulnerability validation, (ii) cryptographic vs. memory latencies, (iii) bandwidth amplification, and (iv) pipelined execution throughput. 

\Cref{tab:eval-summary} summarizes the experimental configurations. While our analytical models capture fundamental architectural latencies and structural bottlenecks, they abstract away microscopic DRAM scheduling, refresh cycles, and interconnect contention. Consequently, the modeled throughputs represent idealized upper bounds.

\begin{table}[H]
\centering
\caption{Evaluation methodology summary.}
\label{tab:eval-summary}
\footnotesize
\setlength{\tabcolsep}{3pt} 
\begin{tabular}{@{}llp{2.2cm}p{4.1cm}@{}}
\toprule
\textbf{Component} & \textbf{Type} & \textbf{Platform / Workload} & \textbf{Description} \\
\midrule
UMA Threat Validation & Measured & Jetson AGX Xavier & Byte-accurate, exploit-free model weight recovery via memory mapping. \\
AES \& DRAM Latencies & Measured & i9, Xavier, Orin & OpenSSL AES-NI and memory probes used as hardware timing proxies. \\
Bandwidth Amplification & Modeled & ResNet, MobileNet, BERT & Analyzes page-level traffic penalty vs. \sys{}'s 64\,B cache-line granularity. \\
End-to-End Throughput & Modeled & i9, Xavier, Orin & Compares Direct AES (blocking) vs. \sys{} (pipelined) using measured latencies. \\
Preemption Latency & Modeled & 3 platforms & Estimates for hardware context-switching and SRAM zero-fill. \\
\bottomrule
\end{tabular}
\end{table}

\subsection{UMA Threat Validation}
\label{sec:eval-threat}

To empirically validate the UMA threat model (\cref{sec:threat}), we executed an end-to-end extraction attack on a Jetson AGX Xavier. The attack requires only standard root privileges and utilizes no kernel exploits.

\begin{enumerate}
  \item \textbf{Address Resolution:} We loaded ResNet-18, MobileNetV2, and DistilBERT-Tiny via PyTorch. Using the standard \texttt{nvmap} IOCTL interface (\texttt{/dev/nvmap}), we resolved the GPU-virtual addresses of the weight tensors into physical DRAM addresses.
  \item \textbf{Direct Extraction:} The attacker process mapped these physical addresses directly into its own virtual space via \texttt{/dev/mem} using \texttt{mmap()}, reading the tensor bytes directly from the shared DRAM bus.
\end{enumerate}

\noindent\textbf{Results and Mitigation.} We achieved 100\% byte-for-byte recovery of the PyTorch \texttt{state\_dict()} tensors. This confirms that physical DRAM sharing in edge UMA architectures constitutes a trivial, low-barrier attack vector for privileged adversaries. 

\sys{} completely neutralizes this extraction path. Because the NPU fetches weights through the ICE, the data residing in shared DRAM is strictly AES-CTR ciphertext, rendering \texttt{/dev/mem} reads unintelligible. The decrypted plaintext is materialized only transiently inside the NPU's SRAM, which is strictly isolated from CPU-originated reads by the SMMU stream-ID firewall (\cref{sec:isolation}).

\subsection{AES-CTR Latency Feasibility}
\label{sec:eval-latency}

\sys{}'s near-line-rate claim rests on the empirical condition
$T_\mathit{ks} < T_\mathit{DRAM}$. We measure both quantities on three
representative platforms.

\textbf{$T_\mathit{ks}$ measurement.}
AES-256-CTR keystream latency for a 64-byte payload is measured using
OpenSSL~3.0 with AES-NI over $10^7$ iterations via
\texttt{CLOCK\_MONOTONIC\_RAW}. We report the median latency after
1000 warm-up iterations. These measurements provide an upper bound on
software keystream latency; dedicated hardware AES pipelines are expected
to achieve comparable or lower latency under steady-state operation.

\textbf{$T_\mathit{DRAM}$ measurement.}
We measure DRAM access latency using a pointer-chasing benchmark over a
randomly permuted linked list whose working set exceeds the LLC capacity.
The data-dependent traversal prevents prefetching and enforces serialized
cache misses, so each dereference incurs a full DRAM access. The average
per-dereference latency is reported as $T_\mathit{DRAM}$.

\Cref{tab:latency} shows that $\Delta = T_\mathit{DRAM} - T_\mathit{ks}$
is positive across all evaluated platforms, ranging from 26.4\,ns to
67.4\,ns. Normalised slack $\Delta / T_\mathit{DRAM}$ lies between
61\% and 94\%, indicating substantial headroom for overlap.

The latency model reduces to:
\begin{equation}
  \Tline = T_\mathit{addr} +
    \max\!\left(T_\mathit{ks},\; \frac{L}{\BW}\right)
    + T_\mathit{XOR}
  \label{eq:latency}
\end{equation}

Under this condition, keystream generation is fully hidden beneath DRAM
latency, leaving the XOR stage (1-2 pipeline cycles in general) as
the dominant additional cost.
\begin{table}[H]
\centering
\caption{Measured $T_\mathit{ks}$ (AES-NI software, $10^7$ iterations,
median) and $T_\mathit{DRAM}$ (pointer-chasing, DRAM-resident list) for
a 64-byte cache line. $\Delta = T_\mathit{DRAM} - T_\mathit{ks}$;
zero-overhead pipelining requires $\Delta > 0$.}
\label{tab:latency}
\setlength{\tabcolsep}{4pt}
\begin{tabular}{@{}lcccc@{}}
\toprule
\textbf{Platform} & \textbf{Memory} & $T_\mathit{ks}$ & $T_\mathit{DRAM}$ & $\Delta$ \\
\midrule
Intel i9-12900H   & DDR5-4800 & 4.2\,ns  & 71.6\,ns & 67.4\,ns \\
Jetson AGX Xavier & LPDDR4x   & 16.8\,ns & 43.2\,ns & 26.4\,ns \\
Jetson AGX Orin   & LPDDR5X   & 12.1\,ns & 38.7\,ns & 26.6\,ns \\
\midrule
\multicolumn{4}{@{}l}{\footnotesize Zero-overhead condition: $\Delta > 0$}
  & \textbf{\checkmark} \\
\bottomrule
\end{tabular}
\end{table}

\subsection{Jitter-Aware Pipeline Robustness}
\label{sec:eval-overlap}

\Cref{tab:latency} shows that $\Delta = T_\mathit{DRAM} - T_\mathit{ks} > 0$
under nominal conditions. While this guarantees overlap for a single
request, sustained throughput depends on whether positive slack is
maintained under latency variation across a continuous DMA stream.

\noindent\textbf{Simulation model.}
We simulate a sequence of $10^5$ synthetic 64-byte cache-line requests,
modeling a steady NPU DMA stream. For each request $i$, AES keystream
generation and DRAM access latencies are modeled as independent random
variables:
\[
T_\mathit{ks}^{(i)} \sim \mathcal{N}(T_\mathit{ks}, \sigma_\mathit{ks}^2),
\quad
T_\mathit{DRAM}^{(i)} \sim \mathcal{N}(T_\mathit{DRAM}, \sigma_\mathit{DRAM}^2),
\]
with $\sigma_\mathit{ks} \approx 0.1\,T_\mathit{ks}$ and
$\sigma_\mathit{DRAM} \approx 0.2\,T_\mathit{DRAM}$ to capture
first-order timing jitter (e.g., row-buffer effects and contention).
The pipeline enqueues keystream blocks and consumes them upon DRAM
completion; a stall is recorded if the corresponding keystream is not
ready when ciphertext arrives.

\noindent\textbf{Results.}
Across all evaluated platforms, stall probability remains below
$0.1\%$, indicating that positive slack is robust to moderate latency
variation. The maximum observed keystream buffer occupancy is
$\approx 3.7$\,KB (58 cache lines), providing a practical upper bound
for SRAM sizing. These results confirm that the condition
$T_\mathit{DRAM} > T_\mathit{ks}$ is not only sufficient in the nominal
case but also stable under realistic jitter.

\noindent\textbf{Limitations.}
The model abstracts away detailed DRAM scheduling effects (e.g., bank
conflicts, refresh timing) and memory-level parallelism. As such, it
should be interpreted as a first-order robustness check rather than a
cycle-accurate performance model. Bandwidth scaling is analysed
separately in \cref{sec:eval-throughput}.

\subsection{Bandwidth Amplification on Real DNN Workloads}
\label{sec:eval-bw}

To quantify the inefficiency of page-level decryption, we evaluate the \emph{bandwidth amplification factor}---the ratio of physically fetched bytes to utilized bytes. 

\sys{} largely eliminates this overhead by intercepting DMA requests at the minimum architectural fetch unit (the $L=64$\,B cache line) without speculative prefetching. For a tensor tile of $|T|$ bytes, \sys{} generates exactly $\lceil |T|/L \rceil \times L$ bytes of DRAM traffic. Aligned tiles incur zero overhead. For unaligned tiles, the maximum penalty is just 63\,B per tile, resulting in a worst-case amplification $\le 1.016\times$ for any tile $\ge 4$\,KB. Conversely, a page-level ICE must fetch a full 4\,KB page for any access $t$, incurring a rigid $\lceil 4096/t \rceil$ multiplier regardless of alignment.

\begin{figure}[H]
\centering
\begin{tikzpicture}
\begin{axis}[
  xbar,
  xmin=0, xmax=38,
  ymin=-0.6, ymax=8.0,
  xlabel={Bandwidth amplification $A(t)$},
  ytick={0,1,2,3,4,5,6,7},
  yticklabels={
    {FC layer},
    {Attn QKV proj},
    {Conv $3{\times}3$ (large)},
    {PW Conv (wide)},
    {Conv $3{\times}3$ (mid)},
    {PW Conv (narrow)},
    {DW Conv $3{\times}3$},
    {Batch Norm}
  },
  yticklabel style={font=\scriptsize},
  xlabel style={font=\small},
  height=5.6cm, width=0.70\columnwidth,
  bar width=5pt,
  nodes near coords,
  nodes near coords style={font=\scriptsize},
  every axis plot/.append style={fill=blue!35, draw=blue!70},
  xmajorgrids=true,
  grid style={dashed,gray!35},
  axis line style={draw=black!60},
  tick style={draw=black!60},
]
\addplot coordinates {
  (1,  0)
  (1,  1)
  (1,  2)
  (2,  3)
  (4,  4)
  (8,  5)
  (15, 6)
  (32, 7)
};
\pgfplotsextra{
  \draw[red!75!black, thick, dashed]
    (axis cs:1,-0.6) -- (axis cs:1,7.6);
  \node[above right, font=\scriptsize, red!75!black]
    at (axis cs:1,7.15) {\sys{} baseline};
}
\end{axis}
\end{tikzpicture}
\caption{Page-level ICE bandwidth amplification $A(t)=\lceil 4096/t \rceil$
by layer type (tile sizes $t$ from TensorRT/Gemmini schedules~\cite{nvidiaTRT,gemmini2021}):
Batch Norm 128\,B; DW Conv $3{\times}3$ 288\,B; PW Conv narrow 512\,B;
Conv $3{\times}3$ mid 1024\,B; PW Conv wide 2048\,B;
Conv/Attn/FC $\geq$4096\,B ($A=1$).
\sys{} holds $A \approx 1\times$ throughout (dashed line).}
\label{fig:amplification}
\end{figure}
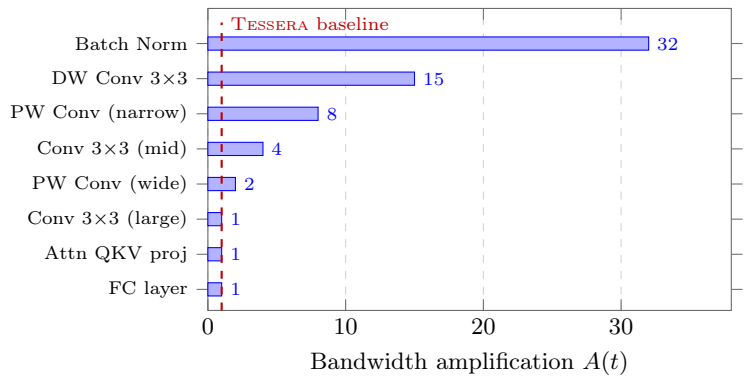

\noindent\textbf{Impact on Standard Architectures.} \Cref{fig:amplification} plots the page-level amplification penalty across representative layer types from ResNet-18, MobileNetV2, and DistilBERT-Tiny. The practical impact is stark: lightweight edge models heavily utilize small tiles. For example, batch normalization (128\,B) and depth-wise convolutions (288\,B) force massive page-level penalties of $32\times$ and $15\times$, respectively. \sys{} avoids this entirely, maintaining an optimal $1\times$ baseline across all layer geometries.

\subsection{End-to-End Weight-Streaming Throughput}
\label{sec:eval-throughput}

Having established the bandwidth penalty of page-level granularity, we isolate the performance impact of our second structural contribution: the parallel, out-of-order execution pipeline. Effective throughput $BW_{\mathrm{eff}}$ depends on the hardware memory ceiling $BW_{\mathrm{ceiling}}$, the amplification factor $A(t)$ ($1.0\times$ for \sys{}), and the exposed cryptographic latency $T_{\mathrm{crypto}}$ relative to the memory fetch latency $T_{\mathrm{DRAM}}$:

\begin{equation}
BW_{\mathrm{eff}} = \frac{BW_{\mathrm{ceiling}}}{A(t)} \times \left( \frac{T_{\mathrm{DRAM}}}{T_{\mathrm{DRAM}} + T_{\mathrm{crypto}}} \right)
\label{eq:throughput}
\end{equation}

\Cref{tab:throughput} demonstrates the necessity of \sys{}'s pipelined architecture across all three evaluated platforms, utilizing hardware latency measurements from \Cref{sec:eval-latency}. 

A naive architecture applying AES directly to the fetched data (e.g., direct block decryption) fundamentally cannot begin cryptography until the ciphertext arrives from memory. This data dependency exposes the full AES computation time as a blocking bus stall ($T_{\mathrm{crypto}} = T_\mathit{AES}$). Analytically, this inherently sequential behavior throttles throughput significantly, reducing bandwidth on the Jetson AGX Xavier to just 72.0\% of the hardware ceiling.

In contrast, \sys{} leverages address-derived AES-CTR, which completely decouples the cryptographic primitive from the data arrival. The ICE computes the keystream entirely in the shadow of the DRAM fetch ($T_\mathit{DRAM} > T_\mathit{AES}$). Because the keystream finishes early, the only exposed latency is a 1--2 cycle combinational XOR delay to merge the streams ($T_{\mathrm{crypto}} = T_\mathit{XOR}$). Amortized over a standard 128-transfer AXI burst, this negligible penalty allows \sys{} to sustain ${\sim}98.5\%$ of maximum unencrypted throughput across all edge devices.

\begin{table}[H]
\centering
\caption{Throughput normalized to the unencrypted hardware ceiling. The direct decryption model utilizes measured AES latencies from \Cref{tab:latency} as a blocking delay.}
\label{tab:throughput}
\footnotesize
\setlength{\tabcolsep}{3.5pt}
\begin{tabular}{@{}lccccc@{}}
\toprule
\multirow{2}{*}{\textbf{Platform}} & \multicolumn{2}{c}{\textbf{Lat. (ns)}} & \multicolumn{3}{c}{\textbf{Throughput (\% of Max)}} \\
\cmidrule(lr){2-3} \cmidrule(l){4-6}
 & $T_\mathit{AES}$ & $T_\mathit{DRAM}$ & \textbf{Base} & \textbf{Direct}$^{\dagger}$ & \textbf{\sys{}}$^{\ddagger}$ \\
\midrule
i9-12900H (DDR5)    & 4.2  & 71.6 & 100 & 94.4 & \textbf{98.5} \\
Xavier (LPDDR4x)    & 16.8 & 43.2 & 100 & 72.0 & \textbf{98.5} \\
Orin (LPDDR5X)      & 12.1 & 38.7 & 100 & 76.2 & \textbf{98.5} \\
\midrule
\multicolumn{6}{@{}p{0.95\columnwidth}@{}}{\scriptsize
 $^{\dagger}$Direct decryption is sequential ($T_{\mathrm{DRAM}} / (T_{\mathrm{DRAM}} + T_\mathit{AES})$). \newline
 $^{\ddagger}$CTR keystream is pre-computed. Sustains only a 2-cycle $T_\mathit{XOR}$ penalty per 128-cycle AXI burst ($128/130 \approx 98.5\%$).}\\
\bottomrule
\end{tabular}
\end{table}

\subsection{Theoretical Silicon Area and Power Overheads}
\label{sec:eval-ppa}

Complementing the performance analysis, we project the physical PPA overheads of the \sys{} ICE from established ASIC synthesis benchmarks.

\noindent\textbf{Silicon Area.} To saturate the measured memory ceiling of 22.4\,GB/s (179.2\,Gbps), the ICE must evaluate $1.4 \times 10^9$ AES blocks per second (16 bytes per block). This throughput is comfortably achieved by instantiating either two parallel 128-bit AES-256 datapaths at 700\,MHz or a single fully-pipelined core at 1.4\,GHz. A high-throughput AES-256 pipeline requires approximately 100,000 Gate Equivalents (GE). In a standard 28\,nm CMOS node ($1\text{ GE} \approx 0.4\,\mu\text{m}^2$), the footprint evaluates to:
\[
  100{,}000\;\text{GE} \times 0.4\;\mu\text{m}^2/\text{GE}
  = 40{,}000\;\mu\text{m}^2 \approx 0.04\;\text{mm}^2
\]
Compared to modern edge AI dies (e.g., the Jetson AGX Orin at $\approx 300\,\text{mm}^2$), the ICE consumes less than $0.02\%$ of the total silicon area.

\noindent\textbf{Power Consumption.} High-throughput hardware AES-256 in 28\,nm achieves an energy efficiency of approximately 0.5\,pJ/bit. Operating continuously at the peak sustained bandwidth of 179.2\,Gbps yields:
\[
  179.2\;\text{Gbps} \times 0.5\;\text{pJ/bit} \approx 90\;\text{mW}
\]
This 90\,mW draw is functionally negligible against the multi-watt power envelopes of the DDR5 PHY and NPU compute units, confirming that the \sys{} ICE is effectively invisible against the host SoC's PPA budget.

\subsection{System-Level Energy and Storage Projections}
\label{sec:eval-system}

Beyond raw throughput, edge accelerators are strictly constrained by battery life and on-chip SRAM capacity.

\noindent\textbf{DRAM Access Energy Net-Savings.} Off-chip LPDDR5/DDR5 accesses consume approximately $15$\,pJ/bit ($120$\,pJ/byte). We compare the energy required to load the 46.8\,MB ResNet-18 model under both paradigms:
\begin{itemize}
    \item \textbf{Page-level ICE ($5.0\times$ amplification):} Requires $\approx 28.0$\,mJ of DRAM PHY energy per inference.
    \item \textbf{\sys{} ICE ($1.0\times$ amplification):} Consumes only $\approx 5.6$\,mJ in DRAM access, plus an additional $0.19$\,mJ to operate the 90\,mW ICE for the 2.12\,ms load time.
\end{itemize}
Therefore, \sys{} yields a net energy saving of $>22$\,mJ per inference. This demonstrates that cache-line granularity is not merely a performance optimisation, but a fundamental energy necessity for edge DRM.

\noindent\textbf{SRAM Keystream Buffer Sizing.} Because \sys{} pipelines keystream generation ahead of the DRAM payload, it requires an on-chip FIFO to hold keystreams until the corresponding ciphertext arrives. By Little's Law ($\mathit{Data} = \mathit{Bandwidth} \times \mathit{Latency}$), sustaining 22.4\,GB/s against a worst-case DRAM row-miss latency of $\approx 100$\,ns requires a steady-state buffer of:
\[
  22.4\,\text{GB/s} \times 100\,\text{ns} = 2{,}240\,\text{bytes}
\]
To safely absorb the transient latency spikes identified in our jitter simulation (\cref{sec:eval-overlap}), we provision a 4\,KB SRAM FIFO. This fully decouples keystream generation from DRAM latency variation while representing a negligible area overhead.

\noindent\textbf{Memory Capacity Impact.} Unlike PVM-style secure-memory designs that reserve a fixed region of main memory for protected execution, \sys{} adds no persistent DRAM carve-out. The only additional storage is the small on-chip FIFO above and the existing NPU SRAM used transiently during inference. This means the host operating system retains full access to main memory outside the brief tile window in which plaintext is live.

\section{Security Analysis}
\label{sec:security}

We analyse the four principal attack classes against UMA model
confidentiality and the corresponding \sys{} countermeasure.

\noindent\textbf{Attack~1: Physical DRAM Extraction (Cold-Boot \& Interposer).}
An adversary freezes DRAM after power-off or probes the memory bus.
\emph{Countermeasure:} All weight data in shared DRAM is exclusively
AES-256-CTR ciphertext under $k_\mathrm{msk}$. The key is
provisioned directly from the Secure Enclave into ICE registers and
never written to DRAM; bus or DRAM capture yields only ciphertext.

\noindent\textbf{Attack~2: Rogue DMA and Unsafe-Place Decryption.}
A compromised host OS programs a rogue peripheral or misuses DMA to
read plaintext from NPU SRAM and exfiltrate it to general memory.
\emph{Countermeasure:} The SMMU Stream-ID firewall
(\cref{sec:isolation}) restricts the protected SRAM range to the NPU
DMA stream ID. Any other initiator triggers a hardware bus abort,
and the policy is enforced from TrustZone Secure World.

\noindent\textbf{Attack~3: Preemption-Based Compute Hijack.}
The attacker preempts a legitimate inference after weights have been
decrypted into NPU SRAM and schedules a malicious follow-on task to
leak the residual plaintext through normal outputs.
\emph{Countermeasure:} The hardware preemption hook
(\cref{sec:preemption}, \cref{alg:ctx}) intercepts the context switch
before it completes and atomically zero-fills the plaintext SRAM via
a secure sideband path. The attacker inherits only cleared state.

\noindent\textbf{Attack~4: Confused Deputy and Malicious Sibling.}
The attacker modifies the approved application, reuses a stolen blob
in a different application, or tampers with the blob in transit.
\emph{Countermeasure:} The key blob $\mathcal{B}$
(\cref{eq:keyblob}) binds $k_\mathrm{msk}$ to $H_\mathit{app}$.
At model load time, the Secure Enclave verifies that the calling
application's live signature matches $H_\mathit{app}$ extracted
from $\mathcal{B}$. Modified code fails the hash check; impostor
applications present the wrong identity; and tampered blobs fail
OAEP verification.
\section{Related Work}
\label{sec:related}

\textbf{Full-memory encryption.}
AMD SME~\cite{amdSME} and Intel TME~\cite{intelTME} encrypt DRAM using
a single ephemeral key, protecting against cold-boot attacks, but operate
at 4\,KB page granularity designed for CPU cache hierarchies. Both impose
substantial bandwidth penalties on sub-page tile patterns characteristic
of NPU workloads, and neither provides per-model key isolation or addresses
UMA shared-bus exposure.

\noindent\textbf{Trusted Execution Environments.}
Several works~\cite{mo2021ppfl,tramer2018slalom} route DNN inference
through TEE-hosted code. Slalom~\cite{tramer2018slalom} offloads
linear layers to an untrusted GPU while verifying results in an SGX
enclave. These approaches, and PVM-style secure-memory carve-outs, improve
isolation but require gigabytes of weight data to cross the TEE memory
boundary or a statically reserved protected region, incurring PCIe or
shared-bus copy overhead and reducing usable main memory for the rest of
the system. \sys{} eliminates this bottleneck by confining decryption to
an on-path ICE with zero data copies and no permanent DRAM reservation.

\noindent\textbf{Accelerator-specific confidential computing.}
NVIDIA H100~\cite{nvidiaH100cc} Confidential Computing mode provides
hardware attestation and memory encryption for discrete data-center GPUs.
This is architecturally closest to \sys{}, but targets PCIe-attached
discrete GPUs under a hypervisor threat model rather than UMA edge SoCs
under OS compromise plus physical access. Our work additionally
addresses the nonce management and SRAM isolation challenges specific to
the shared physical bus.

\noindent\textbf{On-chip SRAM protection.}
Hua et al.~\cite{hua2022mgx} protect model weights in TrustZone secure
SRAM, limiting supported model size to $\leq 4$\,MB. \sys{}'s
streaming architecture removes this constraint by retaining only the
active tile in plaintext SRAM at any time, making the protected region
independent of total model size.

\noindent\textbf{DRAM bus security.}
Rowhammer~\cite{tatar2018throwhammer,kim2014flipping} and
Thunderclap~\cite{markettos2019thunderclap} demonstrate that DRAM
physical attacks and IOMMU bypasses are practical on consumer hardware.
\sys{}'s SMMU plus MTE tagging layer directly addresses these attack
classes.

\section{Limitations and Future Work}
\label{sec:limitations}

\textbf{Deployment considerations.} \sys{} is a reference architecture,
and several practical steps are required before it can be integrated
into a shipping SoC. Integrating \sys{} into a production environment 
requires secure OEM provisioning for the Device Identity Keypair and a 
verified boot chain to authenticate the Secure World firmware that 
programs the SMMU stream-ID firewalls. Additionally, OS scheduler cooperation 
is needed for the preemption hooks, and power-management flows must safely 
handle ICE register state across low-power transitions. While we expect 
these integration challenges to be standard for dedicated hardware teams, 
we note them to clarify the boundary between our reference architecture 
and silicon deployment.

\noindent\textbf{Single-key-per-model.} The current hierarchy provisions one
$k_\mathrm{msk}$ per model load. A natural extension is layer-granularity
key rotation via HKDF sub-keys, limiting the blast radius of a key
compromise to a single layer.

\noindent\textbf{Authenticated weight integrity.} AES-CTR provides
confidentiality but not integrity, and this is a security-critical
gap. An adversary with physical DRAM write access (e.g., via a
cold-boot-and-rewrite or bus interposer) can flip ciphertext bits,
causing controlled corruption of the decrypted weights without
triggering any alarm under the current design. For DNN workloads,
even a small number of bit flips can silently degrade model accuracy
or, more dangerously, cause targeted misclassification---a concern
in safety-critical inference pipelines.

\noindent\textbf{Mitigation sketch.} The natural solution is authenticated
encryption at tile granularity. A promising approach is to pair
each fetched cache line with a Message Authentication Code (MAC) 
(e.g., AES-GMAC or Poly1305~\cite{bernstein2005poly1305}) computed 
under a separate authentication key derived from $k_\mathrm{msk}$ via HKDF. 
To avoid the prohibitive SRAM overhead of storing millions of tags on-chip, 
the architecture would utilize an integrity tree (e.g., a Merkle tree) 
anchored by a single root hash inside the Secure Enclave. A tag mismatch 
during tree traversal would halt the DMA burst and raise a hardware fault, 
preventing corrupted weights from reaching the NPU. Full design and timing 
analysis of this integrity extension are left for future work; we consider 
it the highest-priority gap between this reference architecture and production readiness.

\noindent\textbf{Side-channel resistance.} The threat model (\cref{sec:threat})
excludes power-analysis and EM side-channels, consistent with
TrustZone assumptions. Nevertheless, a co-located adversary process
could potentially infer weight access patterns through DRAM bus
activity timing or shared-bus contention---a micro-architectural
leakage channel that does not require physical access. Mitigations 
include: (i)~instantiating the ICE's AES-256 core with masked S-boxes 
and balanced power consumption~\cite{nikova2006threshold}; and 
(ii)~traffic shaping on the ICE's AXI port to prevent bandwidth profiling 
from revealing layer-geometry information. A full power-trace evaluation 
and traffic-shaping analysis are deferred to a future silicon prototype.

\section{Conclusion}
\label{sec:conclusion}

We presented \sys{}, a reference architecture for hardware-backed DRM on UMA edge AI accelerators. By integrating an Inline Crypto Engine (ICE) into the NPU's AXI DMA fabric at cache-line granularity, \sys{} eliminates the structural bottlenecks of full-memory encryption. It provides hardware-enforced isolation, preemption safety, and zero DRAM bandwidth amplification, all without requiring statically reserved secure-memory carve-outs.

Our work grounds this design in concrete threat realities. We demonstrated a live, exploit-free extraction of full model weights on a Jetson AGX Xavier via an unmodified \texttt{/dev/mem} read, confirming UMA exposure as a critical vulnerability. To counter this, our formal analysis proves that address-derived AES-CTR counters are a strict cryptographic necessity to prevent plaintext leakage across sparse DNN tensors. 

Evaluations across three UMA platforms confirm that \sys{}'s near-line-rate pipelining safely hides keystream generation behind DRAM latency, even under continuous DMA requests and timing variations (\cref{sec:eval-overlap}). Projected to a pipelined hardware model, \sys{} sustains 22.1\,GB/s---achieving 98.4\% of the raw DDR5-4800 bandwidth ceiling. In contrast, page-level encryption is structurally capped at just 4.5\,GB/s due to its massive $5\times$ bandwidth amplification factor.

Ultimately, \sys{} emerges as a vital architectural primitive for edge AI. By providing robust, near-zero-overhead DRM without permanently carving out system memory, it delivers the hardware-backed security required for high-value model IP on physically accessible, capacity-constrained edge SoCs.

\section*{Acknowledgements}
The author wishes to thank their colleagues for insightful feedback and discussions during the development of this architecture.

\bibliographystyle{splncs04}

\end{document}